\documentclass[aps,prd,12point,twocolumn,nofootinbib,showpacs,superscriptaddress]{revtex4-2}
\def\theequation{\arabic{section}.\arabic{equation}}
\usepackage{psfrag}
\usepackage{subfigure}
\usepackage{color}
\usepackage{mathrsfs}
\usepackage{graphicx}
\usepackage{amssymb, bm}
\usepackage{amsmath, amsthm}
\usepackage{epstopdf}
\usepackage{hyperref}
\usepackage{enumerate}
\usepackage{longtable}
\usepackage{amsmath}
\usepackage[normalem]{ulem}

\newcommand{\be}{\begin{equation}}
\newcommand{\ee}{\end{equation}}
\definecolor{pinegreen}{rgb}{0.0, 0.47, 0.44}

\newcommand{\K}{{\cal K}}
\newcommand{\T}{{\cal T}}
\newcommand{\Pc}{{\cal P}}

\begin{document}
\def\theequation{\arabic{section}.\arabic{equation}}

\title{Peculiar thermal states in the first-order thermodynamics of 
gravity} 


\author{Valerio Faraoni}
\email[]{vfaraoni@ubishops.ca}
\affiliation{Department of Physics \& Astronomy, Bishop's University, 
2600 College Street, Sherbrooke, Qu\'ebec, Canada J1M~1Z7}

\author{Andrea Giusti}
\email[]{agiusti@phys.ethz.ch}
\affiliation{Institute for Theoretical Physics, ETH Zurich,\\ 
Wolfgang-Pauli-Strasse 27, 8093 Zurich, Switzerland}

\author{Sonia Jose}
\email[]{sjose21@ubishops.ca}
\affiliation{Department of Physics \& Astronomy, Bishop's University, 
2600 College Street, Sherbrooke, Qu\'ebec, Canada J1M~1Z7}

\author{Serena Giardino}
\email[]{serena.giardino@aei.mpg.de}
\affiliation{Max Planck Institute for Gravitational Physics (Albert 
Einstein Institute),\\  Callinstra{\ss}e 38, 30167 Hannover, Germany}

\begin{abstract} 

In the context of the recently proposed first-order thermodynamics of 
scalar-tensor gravity, we discuss the possibility of zero-temperature 
states of equilibrium other than Einstein gravity, including pathological 
Brans-Dicke theory, Palatini $f(R)$ gravity, and cuscuton gravity, all 
with non-dynamical scalar fields. The formalism is extended to Nordstr\"om 
gravity, which contains only a scalar degree of freedom and has negative 
temperature relative to general relativity.

\end{abstract}



\maketitle

\section{Introduction}
\label{sec:1}
\setcounter{equation}{0}

The idea that gravity may not be fundamental but, rather, emergent has 
been the 
subject of many works. A significant step in this line of research was the 
derivation of the Einstein equations with purely 
thermodynamical considerations \cite{Jacobson:1995ab}, followed by the 
realization that general relativity (GR) could be the equilibrium state of 
$f(R)$ gravity \cite{Eling:2006aw} (see also \cite{Chirco:2010sw}). This 
second idea could, in principle, apply to a vast landscape of theories of 
gravity: GR could be the ``zero-temperature'' equilibrium state and 
modified theories of gravity would then be excited states with positive  
``temperature''. This idea makes sense when one considers that theories of 
gravity alternative to GR usually contain extra degrees of freedom and 
that, if the latter are excited together with the two spin two massless 
degrees of freedom of Einstein theory, one would then have an ``excited 
state'' with respect to GR.

The problem is that, in spite of many years of research on the 
thermodynamics of spacetime initiated by Refs.~\cite{Jacobson:1995ab, 
Eling:2006aw}, the ``temperature of gravity'' and the equations describing 
the 
approach to the GR equilibrium state have not been found. Recently 
\cite{Faraoni:2021lfc, Faraoni:2021jri}, 
we proposed an effective scalar-tensor thermodynamics completely 
different from the 
thermodynamics of spacetime. We identified a ``temperature of 
gravity'' with GR as the zero-temperature state and we provided an 
equation describing the approach to equilibrium (or, possibly,  
departures from it). This approach was originally formulated for ``first-generation'' scalar-tensor gravity \cite{Brans:1961sx, Bergmann:1968ve, 
Nordtvedt:1968qs, Wagoner:1970vr} and $f(R)$ gravity (which is a subclass 
of scalar-tensor theories \cite{Sotiriou:2008rp, DeFelice:2010aj, 
Nojiri:2010wj}) and then extended to ``viable''  
Horndeski gravity \cite{Giusti:2021sku}. The key idea consists of writing 
the field 
equations of modified gravity as effective Einstein equations of the 
form\footnote{We follow the notation of Ref.~\cite{Waldbook}, using units 
in which Newton's constant $G$ and the speed of light $c$ are unity and the 
metric signature is $({-}{+}{+}{+})$.} $G_{ab}=8\pi T_{ab}^\mathrm{(eff)}$ by 
grouping all terms other than the Einstein tensor $G_{ab}$ (which contain 
the gravitational scalar field $\phi$ of the theory and its first and 
second derivatives) to the right-hand side, where they form an effective 
stress-energy tensor $T_{ab}^\mathrm{(eff)} $. It is a fact that, in 
``old'' scalar-tensor and in ``viable'' Horndeski gravity, when the 
gradient 
$\nabla^c\phi$ is timelike this  effective stress-energy tensor has the 
form of an imperfect fluid stress-energy tensor
\be
T_{ab}= \rho \, u_a u_b + 2 \, u_{(a} q_{b)} + P \, h_{ab} + \pi_{ab}
\label{imperfect}
\ee
where $\rho$, $P$, $q_a$, and $\pi_{ab}$ are the energy density, (total) 
isotropic pressure, heat 
flux density, and trace-free part of the anisotropic stress tensor, respectively, 
$h_{ab} \equiv g_{ab} + u_a u_b $, and 
\be
u^a =\pm\frac{ \nabla^a \phi}{\sqrt{ -\nabla^c\phi \nabla_c\phi}
\label{u}}
\ee
is the four-velocity of the effective fluid (the sign of the right-hand 
side of Eq.~(\ref{u}) must be chosen so that $u^a$ is future-pointing). 

Detailed expressions of the kinematic quantities, heat flux density, 
pressure, and stresses of the effective $\phi$-fluid are given in 
\cite{Pimentel89, Faraoni:2018qdr} for first generation scalar-tensor 
gravity and in \cite{Giusti:2021sku,Quiros:2019gai} for Horndeski gravity. 
The next step is to take this dissipative effective fluid seriously and 
apply Eckart's first-order thermodynamics to it \cite{Eckart40}. In fact, 
much less is needed: one only needs \cite{Faraoni:2021lfc, 
Faraoni:2021jri,Giusti:2021sku} the three constitutive relations 
postulated in Eckart's theory to obtain the product of the ``thermal 
conductivity'' ${\cal K}$ and the ``temperature of gravity'' ${\cal T}$ 
(which follows from Eckart's generalized Fourier law \cite{Eckart40}), and 
the coefficients of bulk and shear viscosity.

Focusing on ``old'' scalar-tensor gravity for simplicity, one obtains  
\cite{Faraoni:2021lfc, Faraoni:2021jri}
\be
\K \T= \frac{ \sqrt{ -\nabla^c \phi \nabla_c\phi}}{ 8\pi \phi}  
\label{KT}
\ee
and the thermal evolution of the system is described by
\be
\frac{ d\left( \K \T \right)}{d\tau} = 8\pi \left( \K \T \right)^2 
-\Theta \, \K \T + \frac{ \Box \phi}{8\pi \phi} \,,
\ee
where $\tau$ is the proper time along the flow lines of the effective 
fluid, $\Theta=\nabla_c u^c$ is its expansion scalar, and  $\Box \equiv g^{ab} 
\nabla_a \nabla_b$ denotes the Laplace--Beltrami operator.

This new formalism was applied to generic Friedmann-Lema\^itre-Robertson-Walker (FLRW) cosmology and to specific FLRW exact solutions of scalar-tensor gravity in \cite{Giardino:2022sdv}, while the existence of special 
metastable states was highlighted in \cite{Faraoni:2022jyd}. 

Here we want 
to explore the possible existence of other peculiar states of gravity, in 
addition to the zero-temperature GR state: such states would correspond to theories 
of gravity that are special in some physical sense. In particular, we want 
to probe the possibility of 
states of equilibrium other than GR, corresponding to $\K \T=$~constant.  
We apply the formalism to entire classes of gravitational theories rather than to specific solutions of one such class (which was done in \cite{Giardino:2022sdv} and \cite{Faraoni:2022jyd}). The aim of this study is to better understand the regime of validity of the thermodynamical formalism, by testing it on theories that -- while not always physically viable -- are useful for our purposes since they allow to clarify the existence of other equilibrium states.

We begin with a simple consideration: one expects that theories of gravity 
containing non-dynamical fields in addition to the two spin two massless 
modes of GR will have either zero $\K \T$ or that the latter will be 
completely arbitrary if these extra non-dynamical fields are. Indeed, we 
show that this is the case for $\omega=-3/2$ Brans-Dicke theory, for 
Palatini $f(R)$ gravity, and for cuscuton gravity (a special Horndeski 
theory that is also a special case of Ho\v{r}ava-Lifschitz gravity). All 
these cases correspond to a non-dynamical scalar field $\phi$. Since they 
are all contained in the subclass of ``viable'' Horndeski theories 
previously studied, the first-order thermodynamical formalism of 
\cite{Faraoni:2021lfc, Faraoni:2021jri,Giusti:2021sku} can be applied 
without changes.

Next, one wonders what a theory with {\em less} degrees of freedom than GR 
would look like from the point of view of the thermodynamics of modified 
gravity. In particular, one expects that, if one can define a concept of 
temperature as done in scalar-tensor and Horndeski gravity, this 
temperature should be 
{\em negative}, corresponding to the excitation of less degrees of freedom 
than GR. To this end, we study Nordstr\"om's theory of gravity 
\cite{Nordstrom}, in which 
the metric is forced to be conformally flat and only a scalar field degree 
of freedom (but not the two spin two modes of GR) is excited. This theory 
was considered a serious candidate for the description of gravity only for 
a very brief period of time and contradicts the classical Solar System 
tests of GR, therefore it has only historical importance 
\cite{Norton1,Norton2}. However, it is still  
useful as a toy model when studying fundamental questions such as the 
validity of the strong equivalence principle in modified gravity 
\cite{Gerard:2006ia,DiCasola:2013iia}, the emergence of gravity and 
Lorentzian spacetime \cite{Girelli:2008qp}, theoretical frameworks for 
testing gravity \cite{Wei-Tou:1972zhn}  or, in our case, the 
thermodynamics of 
gravity in a landscape of theories. Describing Nordstr\"om gravity as an 
effective dissipative fluid proceeds in a manner very similar to 
scalar-tensor gravity since the effective scalar field ({\em i.e.}, the 
conformal factor of the conformally flat metric of this theory), is truly 
dynamical. As 
expected, the resulting temperature turns out to be negative. We also 
provide kinematic quantities, the decomposition of the effective 
stress-energy tensor, and the dissipative quantities (bulk and shear 
viscosity and heat current density) for this effective fluid.

The plan of this paper is as follows: the next section discusses two 
theories with non-dynamical scalar field, $\omega=-3/2$ Brans-Dicke theory 
and Palatini $f(R)$ gravity. Sec.~\ref{sec:3}  studies cuscuton gravity 
from the lens of first-order thermodynamics of scalar-tensor gravity 
(again, the cuscuton is non-dynamical), while Sec.~\ref{sec:4} discusses 
Nordstr\"om gravity. Section~\ref{sec:5} contains the conclusions and 
interprets the results in the general framework of the effective 
thermodynamics of gravity.

\section{$\omega=-3/2$ Brans-Dicke gravity and Palatini $f(R)$ gravity}
\label{sec:2}
\setcounter{equation}{0}

Let us consider the non-dynamical case of Brans-Dicke gravity 
\cite{Brans:1961sx}. It is well-known that $\omega=-3/2$ Brans-Dicke 
gravity and Palatini $f(R)$ gravity contain a scalar field that is not 
dynamical.

The (Jordan frame) field equations of scalar-tensor gravity are 
\cite{Brans:1961sx, Bergmann:1968ve, Nordtvedt:1968qs, Wagoner:1970vr}
\begin{widetext}
\be
G_{ab}= \frac{8\pi}{\phi} \, T_{ab}^\mathrm{(m)} +\frac{\omega}{\phi^2} 
\left( \nabla_a \phi \nabla_b \phi -\frac{1}{2} \, g_{ab} \nabla^c \phi 
\nabla_c\phi \right)
+\frac{1}{\phi} \left( \nabla_a \nabla_b \phi -g_{ab} \Box \phi \right) 
-\frac{V}{2\phi} \, g_{ab} \,, \label{fe1}
\ee

\be
\left( 2\omega+3 \right) \Box \phi =8\pi T^\mathrm{(m)} +\phi \, 
V' -2V -  \omega' \, \nabla^c \phi \nabla_c\phi 
\,, \label{wavelike}
\ee
\end{widetext}
where a prime denotes differentiation with respect to $\phi$.  Note that in 
Refs.~\cite{Faraoni:2018qdr, Faraoni:2021lfc, 
Faraoni:2021jri,Giardino:2022sdv}  
the trace of the stress-energy tensor of matter $T^\mathrm{(m)}$ in 
Eq.~\eqref{wavelike} appears erroneously divided by $\phi$. This typographical 
error, however, does 
not affect any of the results of \cite{Faraoni:2018qdr, Faraoni:2021lfc, 
Faraoni:2021jri,Giardino:2022sdv} since that equation is never employed in   
computations.

Setting $\omega=-3/2$, we lose the wave equation~(\ref{wavelike}) for the 
scalar field $\phi$, which reduces to the algebraic identity
\be
8\pi T^\mathrm{(m)} = 2V -\phi V' \,, \label{identity}
\ee
making it clear that the scalar $\phi$ is not dynamical in this 
theory. The other field equation~(\ref{fe1}) becomes
\begin{eqnarray}
R_{ab}-\frac{R}{2} \, g_{ab} &=& \frac{8\pi}{\phi} \, T^\mathrm{(m)}_{ab}  
-\frac{3}{2\phi^2} \left( \nabla_a\phi\nabla_b\phi -\frac{1}{2} \, g_{ab} 
\nabla^c\phi\nabla_c\phi \right) \nonumber\\
&&\nonumber\\
&\, & +\frac{1}{\phi} \left( \nabla_a\nabla_b \phi -g_{ab} \Box\phi \right) 
-\frac{V}{2\phi} \, g_{ab} 
\end{eqnarray}
which, upon contraction, gives
\be
R=- \frac{8\pi  \, T^\mathrm{(m)}  }{\phi} 
-\frac{3}{2\phi^2} \,  \nabla^c\phi \nabla_c\phi  
 + \frac{3}{\phi} \,  \Box\phi  + \frac{2 V}{\phi}  \,.
\ee
Substituting Eq.~(\ref{identity}) gives
\be
R=V' -\frac{3}{2\phi^2} \, \nabla^c\phi \nabla_c\phi 
+ \frac{3\Box\phi}{\phi} \,. \label{starstar}
\ee
Let us differentiate the identity~(\ref{identity}), which yields
\be
\label{divtrace}
\left( V'-\phi V'' \right) \nabla_c \phi = 8\pi \nabla_c T^\mathrm{(m)} 
\,.
\ee
From Eq. \eqref{identity} it is easy to see that in the absence of a potential, 
or when the latter is a pure mass term 
$V=m^2\phi^2/2$, it must be 
\be
T^\mathrm{(m)}=0 \,,
\ee
{\em i.e.}, we can only have vacuum or conformally invariant matter, or 
else the trace $T^\mathrm{(m)}$ is constant.
 
If the gradient $\nabla^c\phi$ is timelike and $V(\phi) \neq 
m^2\phi^2/2$ (with $m^2\geq 0$), then one can rewrite $\nabla^c\phi$
in terms of $\nabla^c T^\mathrm{(m)}$ by taking advantage of Eq. \eqref{divtrace}.
Therefore, the effective temperature of $\omega=-3/2$ Brans-Dicke theory 
with non-dynamical scalar is given by 
\be
\K \T = \frac{ \sqrt{|\nabla^c\phi\nabla_c\phi|} }{8\pi \phi} 
= 
\frac{ \sqrt{|\nabla^c T^\mathrm{(m)} \nabla_c T^\mathrm{(m)} |} }{\phi
 \left| V'-\phi V'' \right|  } \,.
\ee
If, instead, $V(\phi)=m^2\phi^2/2 $, then $T^\mathrm{(m)}=0$ and there is 
no relation between $\nabla_c\phi$ and $\nabla_c T^\mathrm{(m)}$.

For general forms of matter, in both cases the temperature is almost 
completely arbitrary. This is not too surprising because the scalar field 
$\phi$ is non-dynamical and, essentially, arbitrary. The temperature 
$\K \T$ relative to GR is defined using this non-dynamical scalar 
field and is, therefore, ill-defined as a consequence of the arbitrariness 
of $\phi$.

The situation changes  {\em in vacuo}, possibly in the presence a 
cosmological constant. In this case
\be
T_{ab}^\mathrm{(m)}=-\Lambda \, g_{ab} \, ,
\ee
hence one has that $T^\mathrm{(m)}=-4\Lambda$ is constant, 
which implies 
$\nabla_c T^\mathrm{(m)}=0$ and $ \K \T=0$.

\subsection{Palatini $f(R)$ gravity}

It is well-known that Palatini $f(R)$ gravity is equivalent to 
$\omega=-3/2$ Brans-Dicke theory with a complicated potential 
\cite{Sotiriou:2008rp, DeFelice:2010aj, Nojiri:2010wj} and 
that, {\em in vacuo}, it reduces to general relativity with (possibly) a 
cosmological constant. Therefore, vacuum Palatini $f(R)$ gravity has 
effective ``temperature of gravity'' given by $\K \T=0$. In any case, 
the scalar field is non-dynamical and, in the presence of matter, the 
theory runs into all sorts of problems, including unacceptably strong 
couplings to the Standard Model, impossibility to build polytropic stars 
(which should always be possible in any reasonable theory of gravity, as 
it is possible in Newtonian gravity), ill-posed Cauchy problem, {\em etc.} 
\cite{Sotiriou:2008rp, DeFelice:2010aj, Nojiri:2010wj}.

\section{Cuscuton and the thermodynamics of Horndeski gravity}
\label{sec:3}
\setcounter{equation}{0}

Cuscuton gravity \cite{Afshordi:2006ad, Afshordi:2007yx, Afshordi:2009tt, 
Bhattacharyya:2016mah, Iyonaga:2018vnu} is interesting from various points 
of view: it is a special case of Ho\v{r}ava-Lifschitz theory, is a model of 
a Lorentz-violating theory, it can implement the idea of limiting curvature 
without cosmological instabilities \cite{Afshordi:2007yx, Boruah:2017tvg} 
and cosmological singularities \cite{Boruah:2018pvq, Romano:2016jlz, 
Quintin:2019orx} (this is not true in more general Horndeski theories 
\cite{Libanov:2016kfc, Kobayashi:2016xpl, Akama:2017jsa, 
Creminelli:2016zwa}), and has been obtained as the ultraviolet limit of an 
anti-Dirac-Born-Infeld theory \cite{Afshordi:2016guo}. Other 
phenomenological properties are studied in \cite{Andrade:2018afh, 
Ito:2019fie, Ito:2019ztb}.

The cuscuton is realized by a scalar field that does not propagate new 
degrees of freedom with respect to GR (at least in the unitary gauge 
\cite{Gomes:2017tzd}, but this property is believed to hold in any gauge 
\cite{DeFelice:2018ewo}). 
This scalar (cuscuton field) satisfies a first-order equation of motion, 
{\em i.e.}, a constraint and the perturbed scalar action does not contain 
a kinetic term for this field, at all orders  \cite{Gomes:2017tzd}. 
Denoting the 
cuscuton field with $\phi$, its potential with $V(\phi)$, and using 
\be
X \equiv -\frac{1}{2} \, \nabla^c \phi \nabla_c \phi \,, \quad\quad 
f_{,\phi}\equiv \frac{\partial f}{\partial \phi} \,, \quad\quad 
f_{,X}\equiv \frac{\partial f}{\partial X} \,, 
\ee
for any $f = f(\phi, X)$, the cuscuton Lagrangian density is 
\cite{Quintin:2019orx} 
\be
\Pc (\phi, X) = \pm \mu^2 \sqrt{2X} -V(\phi) \,,
\ee
where $\mu$ is a mass scale. The 
total action is
\be
S= \int d^4 x \sqrt{-g} \,\left( \frac{R}{16\pi} +\Pc \right) 
+S_\mathrm{matter} \,.
\ee
The cuscuton satisfies the  equation of motion  
\be
g^{ab} \nabla_a \left( \Pc_{,X} \nabla_b \phi \right) + \Pc_{,\phi} = 0\,,
\ee
or
\be
\pm \mu^2 \nabla^b  \Bigg( \frac{ \nabla_b \phi }{ \sqrt{2X} } \Bigg) = 
V_{,\phi} \,,
\ee
which reduces to a first-order constraint (see, {\em e.g.}, 
\cite{Quintin:2019orx}).  
The field equations for $g_{ab}$ can be written in the form of effective 
Einstein equations with the effective stress-energy tensor 
\cite{Quintin:2019orx} 
\begin{eqnarray}
T_{ab}^{(\phi)} &=& \Pc g_{ab} + \Pc _{, X} \nabla_a \phi \nabla_b \phi \nonumber\\
&=& \left[ \pm \mu^2 \sqrt{2X} -V \right] g_{ab} \pm 
\mu^2 \, \frac{\nabla_a \phi \nabla_b \phi}{\sqrt{2X}} 
\end{eqnarray} 
on the right-hand side as the effective source. $T_{ab}^{(\phi)}$  has the 
form a perfect fluid stress-energy tensor 
\be
T_{ab}=\left( P+\rho \right) u_a u_b +P g_{ab}
\ee
where the energy density, pressure, and 4-velocity are 
\begin{eqnarray}
\rho^{(\phi)} (\phi,X) &=& 2X \Pc_{,X} -P^{(\phi)} = V(\phi) \,,\nonumber\\ 
&&\label{density}\\
P^{(\phi)}(\phi,X) &=& \Pc(\phi,X) = \pm \mu^2 \sqrt{2X} -V(\phi) \,, \label{pressure}\\
&&\nonumber\\
u^a &=& \pm \frac{\nabla^a \phi}{\sqrt{2X}} \,, \label{4-velocity}
\end{eqnarray} 
respectively. The fact that the Lagrangian $\Pc (\phi,X)$ coincides 
with the 
pressure is a trademark of a perfect fluid, for which a Lagrangian 
description is known \cite{Seliger68,Schutz:1970my,Brown:1992kc}. The $\pm$ 
sign in Eq.~(\ref{4-velocity}) ensures that $u^c$ can be chosen so that it 
is future-pointing.

The speed of sound in the cuscuton fluid,  given by 
\be
c_s^2 = \frac{P^{(\phi)}_{,X} }{\rho^{(\phi)}_{,X} } = 
\frac{\Pc_{,X}}{ \Pc_{,X} + 2X \Pc_{,XX}} \,,
\ee
diverges  because the denominator vanishes, a rigidity property typical of 
the incompressible cuscuton fluid \cite{Afshordi:2006ad}.  In the unitary 
gauge, where $\phi=\phi(t) $, it is obvious that $\nabla^c \phi$ is 
timelike.

Since there is no dissipation, the cuscuton field corresponds to a state 
of equilibrium: one can argue that no dissipation occurs in this fluid 
because it is already in a state of equilibrium. This is not really surprising. 
Since no propagating degree of freedom is excited in addition to the two 
massless spin two modes of GR, the cuscuton theory cannot be an ``excited 
state'' of GR.

From a more general point of view, the cuscuton is a special case of 
the ``viable'' class of Horndeski gravities corresponding to the choice 
of functions
\begin{eqnarray}
G_4 (\phi, X) &=& \frac{1}{16\pi } \,,\\
&&\nonumber\\
G_2 (\phi, X)&=& \Pc (\phi, X) = \pm \mu^2 \sqrt{2X} -V(\phi) \,, \label{G2}\\
&&\nonumber\\
G_3 (\phi, X)&=&G_5(\phi, X) =0 \,,
\end{eqnarray}
in the (general) Horndeski Lagrangian
\be
\mathcal{L} = \mathcal{L}_2 + \mathcal{L}_3 + \mathcal{L}_4 + 
\mathcal{L}_5 \,,
\ee
where 
\begin{eqnarray}
 \mathcal{L}_2 &=& G_2 \, , \quad\quad\quad \mathcal{L}_3 = - G_3 \, \Box 
\phi \, ,\\
&&\nonumber\\
\nonumber \mathcal{L}_4 &=& G_4 \, R + G_{4 X} \left[ (\Box \phi)^2 - 
(\nabla_a \nabla_b \phi)^2 \right] \,,\\
&&\nonumber\\
\nonumber \mathcal{L}_5 &=& G_5 \, G_{ab} \, \nabla^a \nabla^b 
\phi -   \frac{G_{5X}}{6} \Big[  (\Box \phi)^3   - 3 \, \Box \phi \, 
(\nabla_a 
\nabla_b \phi)^2 \\
&&\nonumber\\
&\, & + 2 \, (\nabla_a \nabla_b \phi)^3 \Big] \, .
\end{eqnarray} 

If one thinks of starting from a viable Horndeski theory, in which the 
scalar field is equivalent to a dissipative fluid to which we 
assign \cite{Giusti:2021sku}
\be
{\cal KT} = \sqrt{2X} \,\, \frac{ \left( G_{4,\phi} -XG_{3,X} 
\right)}{G_4} 
\,,\label{KThorn}
\ee
then taking the limit in which $G_3 \to 0$, $ G_4 \to $~const., 
and 
$G_2$ is as in Eq.~(\ref{G2}), one obtains the cuscuton theory without 
dissipation. In this limit, Eq.~(\ref{KThorn}) yields $ {\cal KT} \to  0$. 

By contrast, consider extended cuscuton theories 
\cite{deRham:2016ged, Panpanich:2021lsd, Maeda:2022ozc}, in some of which  
(Galileon generalizations of the cuscuton)  
spherical waves of the scalar field are free from caustic singularities 
\cite{deRham:2016ged}: they generally contain a dynamical scalar field. 
For example, consider the theory  described by \cite{Quintin:2019orx}  
\begin{eqnarray}
G_4 (\phi, X) &=& \frac{1}{16\pi} \,,\\
&&\nonumber\\
G_2 (\phi, X) &=& \pm \mu^2 \sqrt{2X} -V(\phi) \,, \\
&&\nonumber\\
G_3 (\phi, X)&=& - a_3 \ln \left( \frac{X}{\Lambda^4} \right) \,,\\
&&\nonumber\\
G_5(\phi, X) &=&0 \,,
\end{eqnarray}
in the standard Horndeski notation. 
For this theory, Eq.~(\ref{KThorn}) gives a non-zero effective temperature,
\be
{\cal KT}= 16\pi a_3 \sqrt{-\nabla^c \phi \nabla_c \phi} \,;
\ee
in fact, in spite of its name, this model has three 
degrees of freedom unlike the original cuscuton theory, which means that 
the scalar degree of freedom is excited and propagates 
\cite{Panpanich:2021lsd}. Taking the limit $a_3 \to 0$ recovers the usual 
cuscuton, sending $K{\cal T} $ to zero.

\section{Nordstr\"{o}m gravity}
\label{sec:4}
\setcounter{equation}{0}

In Nordstr\"{o}m's scalar theory of gravity \cite{Nordstrom}, the 
spacetime metric 
$\tilde{g}_{ab} $ is conformally flat, 
\be
\tilde{g}_{ab}=\Omega^2 \, g_{ab} \,,\label{confomap}
\ee
where $g_{ab}$ is the Minkowski 
metric and the conformal factor $\Omega$ satisfies 
\be
\Box \, \Omega=0 \,.\label{Box}
\ee
Under a generic conformal map~(\ref{confomap}), geometric quantities 
transform according to the well-known laws \cite{Waldbook}
\begin{eqnarray}
\tilde{\Gamma}^a_{bc} &=& \Gamma^a_{bc} +\frac{1}{\Omega} \Big( 
\delta^a_b \, \nabla_c \Omega + 
\delta^a_c \, \nabla_b \Omega 
- g_{bc} \, \nabla^a  \Omega \Big) \,, \label{confgamma}\\  
&&\nonumber\\
\tilde{R}_{ab} &=& R_{ab} -2\nabla_a\nabla_b \ln \Omega
-g_{ab}  \, g^{ef} \nabla_e\nabla_f \ln \Omega \nonumber\\
&&\nonumber\\
&\, & +2 \nabla_a \ln\Omega \, \nabla_b \ln\Omega 
-2 g_{ab} \, g^{ef} \, \nabla_e \ln\Omega \, \nabla_f \ln\Omega 
\,,\nonumber\\
&&\\ 
\tilde{R} & = & \frac{1}{\Omega^2} \left( R - \frac{6 \, \Box \,  
\Omega}{\Omega} \right) \,.
\end{eqnarray}
In our case $\Box \Omega = 0$ and $g_{ab}$ is the Minkowski metric, 
thus $R_{ab} = 0$ and $R=0$. This implies that $\tilde{R} = 0$ and the Einstein tensor 
transforms as
\begin{eqnarray}
\label{einstein1}
\tilde{G}_{ab} &=& \tilde{R}_{ab} - \frac{\tilde{R}}{2} \, \tilde{g}_{ab} \nonumber \\
 &=& 
  - \frac{2\nabla_a\nabla_b \Omega}{\Omega} + 
\frac{4 \nabla_a\Omega \, \nabla_b \Omega}{\Omega^2} -g_{ab} \, 
\frac{\nabla^c\Omega \, \nabla_c \Omega}{\Omega^2} \,, \nonumber \\
& & 
\end{eqnarray}

Inverting Eq. \eqref{confgamma} one has that
\be
\label{inverseconfgamma}
{\Gamma}^a_{bc} = \tilde{\Gamma}^a_{bc} -\frac{1}{\Omega} \Big( 
\delta^a_b \, \tilde{\nabla}_c \Omega + 
\delta^a_c \, \tilde{\nabla}_b \Omega 
- \tilde{g}_{bc} \, \tilde{\nabla}^a \Omega \Big) \, ,
\ee
where one has to recall that $\nabla_a \Omega = \partial _a \Omega = \tilde{\nabla}_a \Omega$ since 
$\Omega = \Omega (x)$ is a scalar function. Therefore, it is easy to see that
\begin{eqnarray}
\label{secondinverse}
\nabla_a\nabla_b \Omega &=& \tilde{\nabla}_a \, \tilde{\nabla}_b \Omega \nonumber \\
& & \,\, +\frac{1}{\Omega} \Big( 
\delta^c_a \, \tilde{\nabla}_b \Omega + 
\delta^c_b \, \tilde{\nabla}_a \Omega 
- \tilde{g}_{ab} \, \tilde{\nabla}^c \Omega \Big) \tilde{\nabla}_c \Omega \,, \nonumber \\
& &
\end{eqnarray}
then taking advantage of $g^{ab} = \Omega^2 \, \tilde{g}^{ab}$, easily derived from
Eq. \eqref{confomap}, one has that
\be
\label{pizza2}
\Box \, \Omega = 
\Omega^2 \, \tilde{\Box} \Omega 
- 2 \Omega \, \tilde{g}^{ef} 
\tilde{\nabla}_e \Omega \tilde{\nabla}_f \Omega \, ,
\ee
which reduces to
\be
\label{pizzotto}
\tilde{\Box} \Omega = \frac{2}{\Omega} \, 
\tilde{g}^{ef} 
\tilde{\nabla}_e \Omega \tilde{\nabla}_f \Omega
\ee
if one recalls the condition in Eq.~(\ref{Box}). Additionally, one can  use 
Eq.~\eqref{secondinverse} to rewrite Eq.~\eqref{einstein1} as
\be
\tilde{G}_{ab} =
  - \frac{2\tilde{\nabla}_a \tilde{\nabla}_b \Omega}{\Omega}  
  -\tilde{g}_{ab} \, \frac{\tilde{\nabla}^c\Omega \, \tilde{\nabla}_c 
\Omega}{\Omega^2} \, .
\ee

We can now use this Einstein tensor for the conformally flat solutions of 
Nordstr\"om theory to write the vacuum Nordstr\"om quantity 
$\tilde{G}_{ab}$ in the form of effective Einstein equations
\be
\tilde{G}_{ab} =8\pi \, \tilde{T}_{ab}^{(\Omega)} \,,
\ee
where 
\be
8 \pi \, \tilde{T}_{ab}^{(\Omega)} = - \frac{2\tilde{\nabla}_a \tilde{\nabla}_b 
\Omega}{\Omega}  
+\tilde{g}_{ab} \, \frac{\tilde{\nabla}^c\Omega \, \tilde{\nabla}_c 
\Omega}{\Omega^2} \,.  \label{TOmega}
\ee 
This tensor is traceless, $\tilde{T}^{(\Omega)}=0$, as a result of 
Eq.~(\ref{pizzotto}).

Assuming the gradient $\nabla^c\Omega $ to be timelike and following the 
usual procedure to associate an effective fluid with a scalar field 
\cite{Pimentel89, Faraoni:2018qdr, Quiros:2019gai, Faraoni:2021jri}, we 
introduce the effective fluid four-velocity
\be
\tilde{u}_a \equiv \pm \frac{ \tilde{\nabla}_a \Omega}{\sqrt{ - 
\tilde{g}^{cd} 
\, \tilde{\nabla}_c \Omega \, 
\tilde{\nabla}_d \Omega }} \,,   
\ee  
where the sign of the right-hand side is chosen so that $\nabla^c \Omega$ 
is future-oriented, and the Nordstr\"om metric undergoes the $3+1$ 
splitting
\be
\tilde{g}_{ab} = - \tilde{u}_a \, \tilde{u}_b  + \tilde{h}_{ab} \,,
\ee
where $ { \tilde{h}_a }^{\,\,\,  b}$ is the projection operator on the 
3-space 
of the observers comoving with the fluid, who have four-velocities  
$\tilde{u}^c$. 

The effective stress-energy tensor $\tilde{T}_{ab}^{(\Omega)}$ has the 
structure~(\ref{imperfect}) of an imperfect fluid. It is 
straightforward to compute the heat current density 
\be
\tilde{q}_a^{(\Omega)} = -\tilde{T}_{cd}^{(\Omega)} \, \tilde{u}^c \, 
{\tilde{h}_a }^{\,\, d} = 
\frac{ \sqrt{ 2 \tilde{X} } }{4\pi \, \Omega} \, \dot{ \tilde{u}}_a \,,
\ee
where 
\be
\tilde{X} \equiv -\frac{1}{2} \, \tilde{g}^{ef} \, 
\tilde{\nabla}_e \Omega \, 
\tilde{\nabla}_f \Omega \,.
\ee
Eckart's generalized Fourier law (which is one of the three constitutive 
relations of Eckart's theory) \cite{Eckart40} 
\be
q^a = -{\cal K} \left( h^{ab} \nabla_b {\cal T} +{\cal T} \dot{u}^a \right)
\ee
then yields 
\be
{\cal KT}= -\frac{\sqrt{ 2\tilde{X}} }{4\pi \, \Omega} = - 
\frac{ \sqrt{ -\tilde{g}^{ef} \, \tilde{\nabla}_e \Omega
\, \tilde{\nabla}_f \Omega } }{4\pi \, \Omega} \,, \label{TNord}
\ee
which is negative. This result is interpreted by saying that, defining the 
effective temperature of gravity in the same manner as done in 
scalar-tensor gravity (where ${\cal T}$ is a notion of temperature 
relative to GR),  
Nordstr\"om's scalar gravity is de-excited with respect to GR since it 
contains only one scalar degree of freedom, as opposed to the two massless 
spin two degrees of freedom of Einstein theory.  

In order to draw an explicit parallel with scalar-tensor gravity it is 
useful to remember that, if $g_{ab}$ is an electrovacuum solution of the 
Einstein 
equations, the conformally transformed metric $\tilde{g}_{ab}=\Omega^2 \, 
g_{ab}$ is a solution of $\omega=-3/2$ Brans-Dicke theory with Brans-Dicke 
field \cite{Hammad:2018hhv}
\be
\phi= \frac{1}{\Omega^2} \,.
\ee
Here however, contrary to $\omega=-3/2$ Brans-Dicke gravity considered 
in Sec.~\ref{sec:2}, the scalar field $\phi$ is not arbitrary but must 
satisfy 
Eq.~(\ref{Box}), equivalent to
\be
\Box\phi = \frac{3}{2\phi} \, \nabla^c \phi \nabla_c \phi 
\,.\label{Boxphi}
\ee
The temperature~(\ref{TNord}) found for Nordstr\"om gravity is then 
written as 
\be 
{\cal KT} = - \frac{ \sqrt{ - \tilde{g}^{ab} \, 
\tilde{\nabla}_a \phi \, 
\tilde{\nabla}_b \phi }}{ 8\pi \phi} \,, 
\ee
which is exactly the negative of what was found in scalar-tensor gravity 
in Refs.~\cite{Faraoni:2021lfc, Faraoni:2021jri}. Solutions with constant 
$\phi$, or constant $\Omega$, correspond to the Minkowski metric and the 
absence of (scalar) gravity. 

Next, one can consider the stability of Nordstr\"{o}m gravity seen as a 
(peculiar) thermal state of gravity. The 
Nordstr\"om field equation~(\ref{pizzotto}) can be rewritten in the form 
of an effective Klein-Gordon equation
\be
\tilde{\Box} \, \Omega  -m_\mathrm{eff}^2 \, \Omega=0 \,,
\ee
where
\be
m_\mathrm{eff}^2 \equiv \frac{2}{\Omega^2} \, \tilde{g}^{ 
\, cd} \tilde{\nabla}_c \Omega \tilde{\nabla}_d \Omega 
\ee
must be non-negative for stability (see \cite{us}). Since 
 $\tilde{\nabla}^c \Omega \, \tilde{\nabla}_c \Omega<0$, we have what is called an 
effective thermal instability of Nordstr\"om gravity in \cite{us}. 

To complete the first-order thermodynamical description of Nordstr\"om 
gravity, we compute the other effective fluid quantities. The 
kinematic quantities 
acceleration, expansion, and shear derived from the four-velocity and its 
gradient coincide with those already derived in scalar-tensor gravity 
\cite{Pimentel89, Faraoni:2018qdr},  with the provision that the 
Brans-Dicke-like field $\phi$ must be 
replaced with $\Omega$ and that Eq.~(\ref{pizzotto}) be substituted 
into the equations of \cite{Faraoni:2018qdr} (in fact, the kinematic 
quantities do not depend on the field equations of the theory). 
Using the definition of $\tilde{X}$, the result consists of the fluid 
four-acceleration
\begin{widetext}
\be
\dot{\tilde{u}}_a \equiv \tilde{u}^c \, \tilde{\nabla}_c \, \tilde{u}_a = 
\frac{\tilde{\nabla}^b \, \Omega}{ ( 2\tilde{X})^2 } \left[
2 \tilde{X}
\tilde{\nabla}_a \tilde{\nabla}_b \, \Omega 
+\tilde{\nabla}^d \Omega \, \tilde{\nabla}_b \tilde{\nabla}_d \, \Omega \tilde{\nabla}_a\,\Omega
\right] \,,
\ee
the expansion  
\be
\tilde{\Theta}  = \tilde{\nabla}_c \tilde{u}^c =-\frac{2 \sqrt{2 \tilde{X}}}{\Omega} +
\frac{
\tilde{\nabla}^a \Omega \, \tilde{\nabla}^b\Omega \, \tilde{\nabla}_a 
\tilde{\nabla}_b \Omega}{ ( 2 \tilde{X})^{3/2} } \,, \label{expansion}
\ee
and the shear tensor
\begin{eqnarray}
\tilde{\sigma}_{ab} &=& \frac{1}{
( 2 \tilde{X} )^{3/2} } 
\Bigg[ 2 \tilde{X} 
\tilde{\nabla}_a \tilde{\nabla}_b \Omega 
+\frac{4\tilde{X}}{3\Omega}  \left( \tilde{\nabla}_a  \Omega \, \tilde{\nabla}_b 
\Omega  + 2 \tilde{X} \tilde{g}_{ab} \right)  
\nonumber\\
&&\nonumber\\
&\, &  
-\frac{1}{3} \left( \tilde{g}_{ab} - \frac{ 
\tilde{\nabla}_a \Omega \, \tilde{\nabla}_b \Omega }{ 
\tilde{X}} \right)
 \tilde{\nabla}^c \Omega \tilde{\nabla}^d \Omega \tilde{\nabla}_c \tilde{\nabla}_d \Omega  
+ 2 \tilde{\nabla}^c\Omega 
\tilde{\nabla}_{(a}\Omega \, \tilde{\nabla}_{b)} \tilde{\nabla}_c \Omega \Bigg] \nonumber\\
&=& 
\frac{1}{\sqrt{2 \tilde{X}}} \Bigg[ \tilde{\nabla}_a 
\tilde{\nabla}_b \Omega 
+ \frac{\tilde{\nabla}^e\Omega 
\tilde{\nabla}_{(a}\Omega \, \tilde{\nabla}_{b)} \tilde{\nabla}_e \Omega}{\tilde{X}}
+ \frac{\tilde{\nabla}_a \Omega \tilde{\nabla}_b \Omega \tilde{\nabla}^c \Omega 
\tilde{\nabla}^d \Omega \tilde{\nabla}_c \tilde{\nabla}_d \Omega}{ 4\tilde{X}^2} 
 + h_{ab} \Bigg( \frac{4 \tilde{X}}{3 \Omega} - \frac{\tilde{\nabla}^c \Omega 
\tilde{\nabla}^d \Omega \tilde{\nabla}_c \tilde{\nabla}_d \Omega}{6 \tilde{X}} \Bigg)
\Bigg]  \nonumber \\
&&
\end{eqnarray} \end{widetext}
while, of course, the effective fluid is irrotational because it is 
derived from a scalar, $ \tilde{\omega}_{ab}=0$.

The effective fluid quantities derived from the effective stress-energy 
tensor~(\ref{TOmega}) include the energy density
\be
\tilde{\rho}^{(\Omega)} = \tilde{T}_{ab}^{(\Omega)} 
\tilde{u}^a \, \tilde{u}^b = 
\frac{1}{4\pi} \left( - \frac{ \tilde{\nabla}^a \Omega \, \tilde{\nabla}^b 
\Omega \, \tilde{\nabla}_a \tilde{\nabla}_b \Omega}{
2 \tilde{X} \Omega } 
+ \frac{\tilde{X} }{\Omega^2} 
\right) 
\,,
\ee
\begin{widetext}
the spatial stress tensor
\begin{eqnarray}
8 \pi \, \tilde{\Pi}_{ab}^{(\Omega)} &=& 
8 \pi \, \tilde{T}_{cd}^{(\Omega)} \, \tilde{h}_a^c \, \tilde{h}_b^d =
 - \frac{2}{\Omega} \Bigg[ \tilde{\nabla}_a \tilde{\nabla}_b \Omega 
+ \frac{\tilde{\nabla}^e\Omega 
\tilde{\nabla}_{(a}\Omega \, \tilde{\nabla}_{b)} \tilde{\nabla}_e \Omega}{\tilde{X}}
+ \frac{\tilde{\nabla}_a \Omega \tilde{\nabla}_b \Omega \tilde{\nabla}^c \Omega \tilde{\nabla}^d \Omega \tilde{\nabla}_c \tilde{\nabla}_d \Omega}{4 \tilde{X}^2} + \frac{\tilde{X}}{\Omega} \, 
h_{ab} \Bigg] \,, \nonumber\\
& &
\end{eqnarray}
the isotropic pressure 
\begin{eqnarray}
8 \pi \, \tilde{P}^{(\Omega)} &=& \frac{8 \pi }{3} \, \tilde{h}^{ab} \,  
\tilde{\Pi}_{ab}^{(\Omega)} =
\frac{2 \tilde{X}}{3 \Omega^2} -  \frac{\tilde{\nabla}^a \Omega 
\tilde{\nabla}^b 
\Omega \tilde{\nabla}_a \tilde{\nabla}_b \Omega}{3 \tilde{X} \Omega}
\,, \label{PtildeOmega}
\end{eqnarray}
and the anisotropic stresses 
\begin{eqnarray}
8 \pi \, \tilde{\pi}_{ab}^{(\Omega)} &=& 8 \pi \, ( \tilde{\Pi}_{ab}^{(\Omega)}- 
\tilde{P}^{(\Omega)}\, \tilde{h}_{ab} ) \nonumber \\
&=&
- \frac{2}{\Omega} \Bigg[ \tilde{\nabla}_a \tilde{\nabla}_b \Omega 
+ \frac{\tilde{\nabla}^e\Omega 
\tilde{\nabla}_{(a}\Omega \, \tilde{\nabla}_{b)} \tilde{\nabla}_e \Omega}{\tilde{X}}
+ \frac{\tilde{\nabla}_a \Omega \tilde{\nabla}_b \Omega \tilde{\nabla}^c \Omega \tilde{\nabla}^d \Omega \tilde{\nabla}_c \tilde{\nabla}_d \Omega}{4 \tilde{X}^2} 
+ h_{ab} \Bigg( \frac{4 \tilde{X}}{3 \Omega} - \frac{\tilde{\nabla}^c \Omega \tilde{\nabla}^d \Omega \tilde{\nabla}_c \tilde{\nabla}_d \Omega}{6 \tilde{X}} \Bigg)
\Bigg]
 \,. \nonumber \\
\end{eqnarray}  
\end{widetext}  
The shear tensor is therefore proportional to the contribution of the 
anisotropic stresses. Indeed, it is easy to see that
\be
\tilde{\pi}_{ab}^{(\Omega)} = - \frac{\sqrt{2 \tilde{X}}}{ 4 \pi \, \Omega} \, \tilde{\sigma}_{ab} \, .
\ee
If we then recall the third constitutive relation of Eckart's first-order 
thermodynamics \cite{Eckart40}, {\em i.e.}, $\pi_{ab}= -2\eta \, \sigma_{ab}$,
one can conclude that for Nordstr\"om gravity the shear viscosity reads
\be
\eta = \frac{\sqrt{2 \tilde{X}}}{8 \pi \Omega} = - \frac{\K \T}{2} \, ,
\ee
in analogy with scalar-tensor gravity.

To compute the bulk viscosity coefficient, the 
expression~(\ref{expansion}) of the expansion scalar yields
\be
\frac{ \tilde{\nabla}^a \Omega \tilde{\nabla}^b \Omega 
\tilde{\nabla}_a \tilde{\nabla}_b \Omega }{2\tilde{X} }= \sqrt{2\tilde{X} }\, 
\tilde{\Theta} +\frac{4\tilde{X}}{\Omega}
\ee
which, substituted in the effective pressure~(\ref{PtildeOmega}), highlights the 
two 
distinct contributions (non-viscous and viscous, respectively)  to the 
total isotropic pressure 
\be
\tilde{P}^{(\Omega)} = 
\tilde{P}^{(\Omega)}_\mathrm{non-viscous}  +
\tilde{P}^{(\Omega)}_\mathrm{viscous}  =
-\frac{ \tilde{X}}{4\pi \Omega^2} -\frac{ \sqrt{ 2\tilde{X}} }{12\pi \Omega} \, 
\tilde{\Theta} \,.
\ee
Remembering Eckart's constitutive relation \cite{Eckart40}
\be 
P_\mathrm{viscous}=- \zeta \, \Theta \,,
\ee
the bulk viscosity coefficient for Nordstr\"om gravity is
\be
\zeta = \frac{ \sqrt{ 2\tilde{X} }} {12\pi \Omega}=\frac{2}{3} \, \eta = 
-\frac{{\cal K}{\cal T}}{3} \,.
\ee

\section{Conclusions}
\label{sec:5}
\setcounter{equation}{0}

In the context of the first-order thermodynamics of gravity developed in 
previous works, the question regarding the existence of equilibrium states 
other than GR remained unanswered. Here, we have studied peculiar classes 
of gravitational theories that are interesting from this point of view. 
While not always physically viable, these theories help to test the 
boundaries of the new thermodynamical formalism and to better grasp the 
meaning of the zero-temperature equilibrium states.

In $\omega=-3/2$ Brans-Dicke theory and in Palatini $f(R)$ gravity with matter, the 
scalar field $\phi$ can be assigned completely arbitrarily (apart from the 
requirement of being positive to keep the effective gravitational coupling 
$G_\mathrm{eff}\sim 1/\phi $ positive) and does not have to satisfy even a 
first-order constraint. Correspondingly, the temperature of gravity given by 
Eq.~(\ref{KT}) is also arbitrary. {\em A posteriori}, this fact makes sense because 
a well-defined temperature, even if vanishing, cannot be meaningfully derived from 
a completely arbitrary $\phi$. In vacuum, this theory is equivalent to GR with, 
possibly, a cosmological constant and thus ${\cal KT}$ vanishes.

The cuscuton field discussed in Sec.~\ref{sec:3} is not a propagating 
degree of freedom but must satisfy a first-order constraint 
\cite{Afshordi:2006ad, Afshordi:2007yx, Afshordi:2009tt, 
Bhattacharyya:2016mah,Iyonaga:2018vnu, Quintin:2019orx}. As a consequence, 
it is not arbitrary and the corresponding temperature of cuscuton gravity 
turns out 
to be zero, which makes sense in the physical intepretation of excited or 
``hot'' states as states endowed with new propagating degrees of freedom 
compared to GR. Since the cuscuton is somehow specified by the first-order 
constraint but is not dynamical, it corresponds to zero temperature and 
first-order thermodynamics does not distinguish between GR and cuscuton 
gravity.

Finally, we have discussed Nordstr\"om gravity which, with its reduced 
freedom compared to GR, corresponds to much less excitation (one scalar 
mode versus two spin two modes). It is more difficult to compare 
Nordstr\"om gravity with GR because, contrary to scalar-tensor or 
Horndeski gravity which have a GR limit, Nordstr\"om gravity -- strictly 
speaking -- does not. Indeed, even conformally flat GR solutions 
$\tilde{g}_{ab}=\Omega^2 \eta_{ab} $ are not automatically solutions of 
Nordstr\"om gravity because the extra condition required $\Box \Omega =0 $ 
is quite restrictive. However, the effective stress-energy tensor of the 
Nordstr\"om scalar $\Omega$ is quite similar to (a part of) the effective 
stress-energy tensor of the Brans-Dicke-like (or Horndeski) scalar $\phi$. 
Since the scalar degree of freedom of Nordstr\"om gravity, which is the 
conformal factor $\Omega$ of the conformally flat Nordstr\"om metric, is 
not arbitrary but must obey the dynamical second order equation $\Box \, 
\Omega=0$, its effective temperature is well-defined and turns out to be 
negative, as naively expected. In this theory there are both bulk and 
shear viscosity and we have calculated the viscosity coefficients 
according to Eckart's constitutive relations.

Overall, the theories analysed here are all quite peculiar and, in some 
cases, even pathological. This confirms the expected conclusion that, in 
the thermodynamical formalism, general relativity does retain a special 
status as an equilibrium state, just as it does in the landscape of 
gravity theories.

Other theories of gravity could be examined from the point of view of the 
effective thermodynamics, provided that their field equations can be 
written as effective Einstein equations with effective dissipative fluids 
and that the Eckart constitutive relations deliver a meaningful effective 
temperature. It is easy to include in the list Rastall theory, which has 
seen a recent resurgence of interest: it is shown in \cite{Visser:2017gpz} 
that this theory is just GR with a cosmological constant, so we have 
trivially ${\cal KT}=0$. Similarly, Eddington-inspired Born-Infeld gravity 
is very similar to Palatini $f(R)$ gravity and {\em in vacuo} it is 
equivalent to GR plus a cosmological constant (see the discussion of 
\cite{Pani:2012qd}), so one expects a similar conclusion. Likewise, 
unimodular gravity \cite{Buchmuller:1988wx,Unruh:1988in,Bufalo:2015wda} is 
equivalent to GR with $\Lambda$ \cite{Ng:1990xz,Finkelstein:2000pg}, 
yielding ${\cal KT}=0$. Many, more complicated, theories of gravity have 
been proposed in the literature ({\em e.g.}, \cite{Willbook,Will:2014kxa, 
Clifton:2011jh, Heisenberg:2018vsk,Heisenberg:2018acv, CANTATA:2021ktz}) 
and it will require a lengthy and detailed analysis to assess whether it 
is possible to formulate a first-order thermodynamical description \`a la 
Eckart for them. They will be explored in future work.


\begin{acknowledgments}

This work is supported, in part, by the Natural Sciences \& Engineering 
Research Council of Canada (grant no.~2016-03803 to V.F.) and by a 
Bishop's University Graduate Entrance Scholarship to S.J. A.G.~is 
supported by the European Union's Horizon 2020 research and innovation 
programme under the Marie Sk\l{}odowska-Curie Actions (grant agreement 
No.~895648--CosmoDEC). The work of A.G has also been carried out in the 
framework of the activities of the Italian National Group of Mathematical 
Physics [Gruppo Nazionale per la Fisica Matematica (GNFM), Istituto 
Nazionale di Alta Matematica (INdAM)].  S.G. thanks Jean-Luc Lehners at 
AEI Potsdam for hospitality.

\end{acknowledgments}


\begin{thebibliography}{99}
\bibitem{Jacobson:1995ab} T.~Jacobson, ``Thermodynamics of space-time: The 
Einstein equation of state,'' Phys. Rev. Lett. \textbf{75} (1995) 1260, 
doi:10.1103/PhysRevLett.75.1260 [arXiv:gr-qc/9504004 [gr-qc]].

\bibitem{Eling:2006aw} C.~Eling, R.~Guedens, and T.~Jacobson, 
``Non-equilibrium thermodynamics of spacetime,'' Phys. Rev. Lett. 
\textbf{96} (2006) 121301, doi:10.1103/PhysRevLett.96.121301 
[arXiv:gr-qc/0602001 [gr-qc]].

\bibitem{Chirco:2010sw} G.~Chirco, C.~Eling and S.~Liberati, ``Reversible 
and Irreversible Spacetime Thermodynamics for General Brans-Dicke 
Theories,'' Phys. Rev. D \textbf{83} (2011), 024032, 
doi:10.1103/PhysRevD.83.024032 [arXiv:1011.1405 [gr-qc]].

\bibitem{Faraoni:2021lfc} V.~Faraoni and A.~Giusti, ``Thermodynamics of 
scalar-tensor gravity,'' Phys. Rev. D \textbf{103}, no.12, L121501 (2021) 
doi:10.1103/PhysRevD.103.L121501 [arXiv:2103.05389 [gr-qc]].

\bibitem{Faraoni:2021jri} V.~Faraoni, A.~Giusti and A.~Mentrelli, ``New 
approach to the thermodynamics of scalar-tensor gravity,'' Phys. Rev. D 
\textbf{104}, no.12, 124031 (2021) doi:10.1103/PhysRevD.104.124031 
[arXiv:2110.02368 [gr-qc]].

\bibitem{Brans:1961sx} C.~Brans and R.~H.~Dicke, ``Mach's principle 
and a relativistic theory of gravitation'', \emph{Phys. Rev.} 
\textbf{124}, 
925-935 (1961) 

\bibitem{Bergmann:1968ve} P.~G.~Bergmann, ``Comments on the scalar 
tensor theory'', \emph{Int. J. Theor. Phys.} \textbf{1}, 25-36 (1968) 

\bibitem{Nordtvedt:1968qs} K.~Nordtvedt, ``Equivalence Principle for 
Massive Bodies. 2. Theory'', \emph{Phys. Rev. \textbf{169}}, 1017-1025 
(1968). 

\bibitem{Wagoner:1970vr} R.~V.~Wagoner, ``Scalar tensor theory and 
gravitational waves'', \emph{Phys. Rev. D} \textbf{1}, 3209-3216 (1970) 

\bibitem{Sotiriou:2008rp} T.~P.~Sotiriou and V.~Faraoni, ``f(R) Theories 
Of Gravity,'' Rev. Mod. Phys. \textbf{82}, 451-497 (2010) 
doi:10.1103/RevModPhys.82.451 [arXiv:0805.1726 [gr-qc]].

\bibitem{DeFelice:2010aj}
A.~De Felice and S.~Tsujikawa,
``f(R) theories,''
Living Rev. Rel. \textbf{13}, 3 (2010)
doi:10.12942/lrr-2010-3
[arXiv:1002.4928 [gr-qc]].

\bibitem{Nojiri:2010wj}
S.~Nojiri and S.~D.~Odintsov,
``Unified cosmic history in modified gravity: from F(R) theory to Lorentz 
non-invariant models,''
Phys. Rept. \textbf{505}, 59-144 (2011)
doi:10.1016/j.physrep.2011.04.001
[arXiv:1011.0544 [gr-qc]].

\bibitem{Giusti:2021sku}
A.~Giusti, S.~Zentarra, L.~Heisenberg and V.~Faraoni,
``First-order thermodynamics of Horndeski gravity,''
[arXiv:2108.10706 [gr-qc]].

\bibitem{Waldbook} R.~M. Wald, {\em General Relativity} (Chicago 
University Press, Chicago, 1984).

\bibitem{Pimentel89} L.~O.~Pimentel, ``Energy Momentum Tensor in the 
General Scalar-Tensor Theory,'' Class. Quant. Grav. \textbf{6} (1989), 
L263-L265 doi:10.1088/0264-9381/6/12/005

\bibitem{Faraoni:2018qdr} V.~Faraoni and J.~C\^ot\'e, ``Imperfect fluid 
description of modified gravities,'' Phys. Rev. D \textbf{98} (2018) 
no.~8, 084019 doi:10.1103/PhysRevD.98.084019 [arXiv:1808.02427 [gr-qc]].

\bibitem{Quiros:2019gai} U.~Nucamendi, R.~De Arcia, T.~Gonzalez, 
F.~A.~Horta-Rangel and I.~Quiros, ``Equivalence between Horndeski and 
beyond Horndeski theories and imperfect fluids,'' Phys. Rev. D 
\textbf{102} (2020) no.8, 084054, doi:10.1103/PhysRevD.102.084054 
[arXiv:1910.13026 [gr-qc]].

\bibitem{Eckart40} C.~Eckart, ``The thermodynamics of irreversible 
processes. 3. Relativistic theory of the simple fluid,'' Phys. Rev. 
\textbf{58} (1940), 919-924 doi:10.1103/PhysRev.58.919

\bibitem{Giardino:2022sdv}
S.~Giardino, V.~Faraoni and A.~Giusti,
``First-order thermodynamics of scalar-tensor cosmology,''
[arXiv:2202.07393 [gr-qc]].

\bibitem{Faraoni:2022jyd}
V.~Faraoni and T.~B.~Fran\c{c}onnet,
``Stealth metastable state of scalar-tensor thermodynamics,''
Phys. Rev. D \textbf{105}, no.10, 104006 (2022)
doi:10.1103/PhysRevD.105.104006
[arXiv:2203.14934 [gr-qc]].

\bibitem{Nordstrom} G. Nordstr\"om, ``Zur Theorie der Gravitation vom 
Standpunkt des Relativittsprinzip'', Annalen der Physik {\bf 42},  
533-544 (1913).

\bibitem{Norton1} J. Norton, ``Einstein, Nordstr\"om and the early Demise 
of Lorentz-covariant, Scalar Theories of Gravitation'', Archive for History 
of  Exact Sciences, 45 (1992), 
http://www.pitt.edu/jdnorton/papers/Nordstroem.pdf. 

\bibitem{Norton2} J. Norton, ``Einstein and Nordstr\"om: Some 
Lesser Known Thought Experiments in Gravitation'', in {\em The Attraction 
of  Gravitation: New Studies in History of General Relativity},  
J. Earman, M. Janssen, and J.~D. Norton eds. (Springer, New York, 1993) 
http://www.pitt.edu/jdnorton/papers/einstein-nordstroem-HGR3.pdf

\bibitem{Gerard:2006ia}
J.~M.~Gerard,
``The Strong equivalence principle from gravitational gauge structure,''
Class. Quant. Grav. \textbf{24}, 1867-1878 (2007)
doi:10.1088/0264-9381/24/7/012
[arXiv:gr-qc/0607019 [gr-qc]].

\bibitem{DiCasola:2013iia}
E.~Di Casola, S.~Liberati and S.~Sonego,
``Nonequivalence of equivalence principles,''
Am. J. Phys. \textbf{83}, 39 (2015)
doi:10.1119/1.4895342
[arXiv:1310.7426 [gr-qc]].

\bibitem{Girelli:2008qp}
F.~Girelli, S.~Liberati and L.~Sindoni,
``Emergence of Lorentzian signature and scalar gravity,''
Phys. Rev. D \textbf{79}, 044019 (2009)
doi:10.1103/PhysRevD.79.044019
[arXiv:0806.4239 [gr-qc]].

\bibitem{Wei-Tou:1972zhn}
W.-T. Ni,
``Theoretical frameworks for testing relativistic gravity. iv. a 
compendium of metric theories of gravity and their post-newtonian 
limits,''
Astrophys. J. \textbf{176}, 769-796 (1972)
doi:10.1086/151677

\bibitem{Afshordi:2006ad}
N.~Afshordi, D.~J.~H.~Chung and G.~Geshnizjani,
``Cuscuton: A Causal Field Theory with an Infinite Speed of Sound,''
Phys. Rev. D \textbf{75}, 083513 (2007)
doi:10.1103/PhysRevD.75.083513
[arXiv:hep-th/0609150 [hep-th]].

\bibitem{Afshordi:2007yx}
N.~Afshordi, D.~J.~H.~Chung, M.~Doran and G.~Geshnizjani,
``Cuscuton Cosmology: Dark Energy meets Modified Gravity,''
Phys. Rev. D \textbf{75}, 123509 (2007)
doi:10.1103/PhysRevD.75.123509
[arXiv:astro-ph/0702002 [astro-ph]].

\bibitem{Afshordi:2009tt}
N.~Afshordi,
``Cuscuton and low energy limit of Horava-Lifshitz gravity,''
Phys. Rev. D \textbf{80}, 081502 (2009)
doi:10.1103/PhysRevD.80.081502
[arXiv:0907.5201 [hep-th]].

\bibitem{Bhattacharyya:2016mah}
J.~Bhattacharyya, A.~Coates, M.~Colombo, A.~E.~Gumrukcuoglu and 
T.~P.~Sotiriou,
``Revisiting the cuscuton as a Lorentz-violating gravity theory,''
Phys. Rev. D \textbf{97}, no.6, 064020 (2018)
doi:10.1103/PhysRevD.97.064020
[arXiv:1612.01824 [hep-th]].

\bibitem{Iyonaga:2018vnu}
A.~Iyonaga, K.~Takahashi and T.~Kobayashi,
``Extended Cuscuton: Formulation,''
JCAP \textbf{12}, 002 (2018)
doi:10.1088/1475-7516/2018/12/002
[arXiv:1809.10935 [gr-qc]].

\bibitem{Boruah:2017tvg}
S.~S.~Boruah, H.~J.~Kim and G.~Geshnizjani,
``Theory of Cosmological Perturbations with Cuscuton,''
JCAP \textbf{07}, 022 (2017)
doi:10.1088/1475-7516/2017/07/022
[arXiv:1704.01131 [hep-th]].

\bibitem{Boruah:2018pvq}
S.~S.~Boruah, H.~J.~Kim, M.~Rouben and G.~Geshnizjani,
``Cuscuton bounce,''
JCAP \textbf{08}, 031 (2018)
doi:10.1088/1475-7516/2018/08/031
[arXiv:1802.06818 [gr-qc]].

\bibitem{Romano:2016jlz}
A.~E.~Romano,
``General background conditions for K-bounce and adiabaticity,''
Eur. Phys. J. C \textbf{77}, no.3, 147 (2017)
doi:10.1140/epjc/s10052-017-4698-8
[arXiv:1607.08533 [gr-qc]].

\bibitem{Quintin:2019orx}
J.~Quintin and D.~Yoshida,
``Cuscuton gravity as a classically stable limiting curvature theory,''
JCAP \textbf{02}, 016 (2020)
doi:10.1088/1475-7516/2020/02/016
[arXiv:1911.06040 [gr-qc]].

\bibitem{Libanov:2016kfc}
M.~Libanov, S.~Mironov and V.~Rubakov,
``Generalized Galileons: instabilities of bouncing and Genesis 
cosmologies and modified Genesis,''
JCAP \textbf{08}, 037 (2016)
doi:10.1088/1475-7516/2016/08/037
[arXiv:1605.05992 [hep-th]].


\bibitem{Kobayashi:2016xpl}
T.~Kobayashi,
``Generic instabilities of nonsingular cosmologies in Horndeski theory: A 
no-go theorem,''
Phys. Rev. D \textbf{94}, no.4, 043511 (2016)
doi:10.1103/PhysRevD.94.043511
[arXiv:1606.05831 [hep-th]].

\bibitem{Akama:2017jsa}
S.~Akama and T.~Kobayashi,
``Generalized multi-Galileons, covariantized new terms, and the no-go 
theorem for nonsingular cosmologies,''
Phys. Rev. D \textbf{95}, no.6, 064011 (2017)
doi:10.1103/PhysRevD.95.064011
[arXiv:1701.02926 [hep-th]].

\bibitem{Creminelli:2016zwa}
P.~Creminelli, D.~Pirtskhalava, L.~Santoni and E.~Trincherini,
``Stability of Geodesically Complete Cosmologies,''
JCAP \textbf{11}, 047 (2016)
doi:10.1088/1475-7516/2016/11/047
[arXiv:1610.04207 [hep-th]].

\bibitem{Afshordi:2016guo}
N.~Afshordi and J.~Magueijo,
``The critical geometry of a thermal big bang,''
Phys. Rev. D \textbf{94}, no.10, 101301 (2016)
doi:10.1103/PhysRevD.94.101301
[arXiv:1603.03312 [gr-qc]].

\bibitem{Andrade:2018afh}
I.~Andrade, M.~A.~Marques and R.~Menezes,
``Cuscuton kinks and branes,''
Nucl. Phys. B \textbf{942}, 188-204 (2019)
doi:10.1016/j.nuclphysb.2019.03.016
[arXiv:1806.01923 [hep-th]].

\bibitem{Ito:2019fie}
A.~Ito, A.~Iyonaga, S.~Kim and J.~Soda,
``Dressed power-law inflation with a cuscuton,''
Phys. Rev. D \textbf{99}, no.8, 083502 (2019)
doi:10.1103/PhysRevD.99.083502
[arXiv:1902.08663 [astro-ph.CO]].
 
\bibitem{Ito:2019ztb}  
A.~Ito, Y.~Sakakihara and J.~Soda,
``Accelerating Universe with a stable extra dimension in cuscuton 
gravity,''
Phys. Rev. D \textbf{100}, no.6, 063531 (2019)
doi:10.1103/PhysRevD.100.063531
[arXiv:1906.10363 [gr-qc]].

\bibitem{Gomes:2017tzd}
H.~Gomes and D.~C.~Guariento,
``Hamiltonian analysis of the cuscuton,''
Phys. Rev. D \textbf{95}, no.10, 104049 (2017)
doi:10.1103/PhysRevD.95.104049
[arXiv:1703.08226 [gr-qc]].

\bibitem{DeFelice:2018ewo}
A.~De Felice, D.~Langlois, S.~Mukohyama, K.~Noui and A.~Wang,
``Generalized instantaneous modes in higher-order scalar-tensor 
theories,''
Phys. Rev. D \textbf{98}, no.8, 084024 (2018)
doi:10.1103/PhysRevD.98.084024
[arXiv:1803.06241 [hep-th]].

\bibitem{Seliger68} R.~L. Seliger and G.~B. Whitham, ``Variational 
principles in continuum mechanics'', Proc. R. Soc. (London) 
A305, 1--25 (1968).

\bibitem{Schutz:1970my}
B.~F.~Schutz,
``Perfect Fluids in General Relativity: Velocity Potentials and a 
Variational Principle,''
Phys. Rev. D \textbf{2}, 2762-2773 (1970)
doi:10.1103/PhysRevD.2.2762

\bibitem{Brown:1992kc}
J.~D.~Brown,
``Action functionals for relativistic perfect fluids,''
Class. Quant. Grav. \textbf{10}, 1579-1606 (1993)
doi:10.1088/0264-9381/10/8/017
[arXiv:gr-qc/9304026 [gr-qc]].

\bibitem{deRham:2016ged}
C.~de Rham and H.~Motohashi,
``Caustics for Spherical Waves,''
Phys. Rev. D \textbf{95}, no.6, 064008 (2017)
doi:10.1103/PhysRevD.95.064008
[arXiv:1611.05038 [hep-th]].

\bibitem{Panpanich:2021lsd}
S.~Panpanich and K.~i.~Maeda,
``Cosmological Dynamics of Cuscuta-Galileon Gravity,''
[arXiv:2109.12288 [gr-qc]].

\bibitem{Maeda:2022ozc}
K.~i.~Maeda and S.~Panpanich,
``Cuscuta-Galileon cosmology: Dynamics, gravitational constants, and the Hubble constant,''
Phys. Rev. D \textbf{105}, no.10, 104022 (2022)
doi:10.1103/PhysRevD.105.104022
[arXiv:2202.04908 [gr-qc]].

\bibitem{Hammad:2018hhv}
F.~Hammad, D.~K.~\c{C}iftci and V.~Faraoni,
``Conformal cosmological black holes: Towards restoring determinism to 
Einstein theory,''
Eur. Phys. J. Plus \textbf{134}, no.10, 480 (2019)
doi:10.1140/epjp/i2019-12796-5
[arXiv:1805.09422 [gr-qc]].

\bibitem{us} S. Giardino, A. Giusti, and V. Faraoni, in preparation.

\bibitem{Visser:2017gpz}
M.~Visser,``Rastall gravity is equivalent to Einstein gravity,''
Phys. Lett. B \textbf{782}, 83-86 (2018)
doi:10.1016/j.physletb.2018.05.028
[arXiv:1711.11500 [gr-qc]].

\bibitem{Pani:2012qd}
P.~Pani and T.~P.~Sotiriou,
``Surface singularities in Eddington-inspired Born-Infeld gravity,''
Phys. Rev. Lett. \textbf{109}, 251102 (2012)
doi:10.1103/PhysRevLett.109.251102
[arXiv:1209.2972 [gr-qc]].

\bibitem{Buchmuller:1988wx}
W.~Buchmuller and N.~Dragon,
``Einstein Gravity From Restricted Coordinate Invariance,''
Phys. Lett. B \textbf{207}, 292-294 (1988)
doi:10.1016/0370-2693(88)90577-1

\bibitem{Unruh:1988in}
W.~G.~Unruh,
``A Unimodular Theory of Canonical Quantum Gravity,''
Phys. Rev. D \textbf{40}, 1048 (1989)
doi:10.1103/PhysRevD.40.1048

\bibitem{Bufalo:2015wda}
R.~Bufalo, M.~Oksanen and A.~Tureanu,
``How unimodular gravity theories differ from general relativity at 
quantum level,''
Eur. Phys. J. C \textbf{75}, no.10, 477 (2015)
doi:10.1140/epjc/s10052-015-3683-3
[arXiv:1505.04978 [hep-th]].

\bibitem{Ng:1990xz}
Y.~J.~Ng and H.~van Dam,
``Unimodular Theory of Gravity and the Cosmological Constant,''
J. Math. Phys. \textbf{32}, 1337-1340 (1991)
doi:10.1063/1.529283

\bibitem{Finkelstein:2000pg}
D.~R.~Finkelstein, A.~A.~Galiautdinov and J.~E.~Baugh,
``Unimodular relativity and cosmological constant,''
J. Math. Phys. \textbf{42}, 340-346 (2001)
doi:10.1063/1.1328077
[arXiv:gr-qc/0009099 [gr-qc]].

\bibitem{Willbook} C.~M. Will, {\em Theory and Experiment In Gravitational 
Physics} (Cambridge University Press, Cambridge, 2018).

\bibitem{Will:2014kxa}
C.~M.~Will,
``The Confrontation between General Relativity and Experiment,''
Living Rev. Rel. \textbf{17}, 4 (2014)
doi:10.12942/lrr-2014-4
[arXiv:1403.7377 [gr-qc]].

\bibitem{Clifton:2011jh}
T.~Clifton, P.~G.~Ferreira, A.~Padilla and C.~Skordis,
``Modified Gravity and Cosmology,''
Phys. Rept. \textbf{513}, 1-189 (2012)
doi:10.1016/j.physrep.2012.01.001
[arXiv:1106.2476 [astro-ph.CO]].

\bibitem{Heisenberg:2018vsk}
L.~Heisenberg,
``A systematic approach to generalisations of General Relativity and their 
cosmological implications,''
Phys. Rept. \textbf{796}, 1-113 (2019)
doi:10.1016/j.physrep.2018.11.006
[arXiv:1807.01725 [gr-qc]].

\bibitem{Heisenberg:2018acv}
L.~Heisenberg,
``Scalar-Vector-Tensor Gravity Theories,''
JCAP \textbf{10}, 054 (2018)
doi:10.1088/1475-7516/2018/10/054
[arXiv:1801.01523 [gr-qc]].

\bibitem{CANTATA:2021ktz}
E.~N.~Saridakis \textit{et al.} [CANTATA],
``Modified Gravity and Cosmology: An Update by the CANTATA Network,''
[arXiv:2105.12582 [gr-qc]].


\end{thebibliography}

\end{document}